\documentclass[12pt]{article}
\usepackage{amssymb}
\usepackage{graphicx}

\ifx\pdfoutput\undefined
    \relax
\else
    \usepackage{epstopdf}
\fi
\makeatletter
\def\numberbysection{\@addtoreset{equation}{section}
 	\def\theequation{\thesection.\arabic{equation}}}
\makeatother

\numberbysection

\newcommand{\be}{\begin{eqnarray}}
\newcommand{\ee}{\end{eqnarray}}
\newcommand{\non}{\nonumber}
\newcommand{\tr}{\mathop{\rm tr}\nolimits}

\newcommand{\csch}{\mathop{\rm csch}\nolimits}

\newcommand{\af}{\ensuremath{\mathsf{a}}}
\newcommand{\Af}{\ensuremath{\mathsf{A}}}
\newcommand{\bff}{\ensuremath{\mathsf{b}}}
\newcommand{\Bf}{\ensuremath{\mathsf{B}}}

\newcommand{\Pf}{\ensuremath{\mathsf{P}}}

\def\ep{\epsilon}

\def\La{\Lambda}

\newcommand{\beq}{\begin{equation}}
\newcommand{\eeq}{\end{equation}}
\newcommand{\bea}{\begin{eqnarray*}}
\newcommand{\eea}{\end{eqnarray*}}
\newcommand{\beqa}{\begin{eqnarray}}
\newcommand{\eeqa}{\end{eqnarray}}
\begin{document}

\begin{titlepage}
\strut\hfill
\vspace{.5in}
\begin{center}

\LARGE A note on a boundary sine-Gordon model\\
\LARGE at the free-Fermion point\\[1.0in]
\large Rajan Murgan\footnote{email:rmurgan@svsu.edu}\\[0.8in]
\large Department of Physics,\\ 
\large Saginaw Valley State University,\\ 
\large 7400 Bay Road University Center,  MI 48710 USA\\

\end{center}

\vspace{.5in}

\begin{abstract}
We investigate the free-Fermion point of a boundary sine-Gordon model with nondiagonal boundary interactions for the ground state using auxiliary functions obtained from $T-Q$ equations of a 
corresponding inhomogeneous open spin-$\frac{1}{2}$ XXZ chain with nondiagonal boundary terms. In particular, we obtain the Casimir energy. 
Our result for the Casimir energy is shown to agree with the result from the TBA approach. The analytical result for the effective central charge in the 
ultraviolet (UV) limit is also verified from the plots of effective central charge for intermediate values of volume. 
\end{abstract}
\end{titlepage}

\setcounter{footnote}{0}

\section{Introduction}\label{sec:intro}

Open spin-$\frac{1}{2}$ XXZ quantum spin chains and boundary sine-Gordon (bsG) models have applications in statistical mechanics and condensed matter physics. 
These models have thus been widely studied over the years \cite{Ga}-\cite{Murgeven}. 
Studies conducted have covered various aspects of these models, with much of them focusing on the nondiagonal boundary interactions such as in \cite{Sk,dVGR,GZ} and 
\cite {nondiagonal}-\cite{Murgeven}. In particular, progress made in finding solutions of the open spin-$\frac{1}{2}$ XXZ quantum spin chain with nondiagonal boundary terms 
such as that given in \cite{Nep, Nepnondiag, Nepnondiag2, Mureven1} has motivated further studies of these models \cite{AN, ABNPT, Murgeven}.

Much work has also been done on bsG models at the free-Fermion point (where the lattice bulk parameter $\mu$ takes the value $\mu = \pi/2$), such as in \cite{AKL, LMSS, AN}. Motivated by 
these results, in this paper, we revisit the bsG model 
at the free-Fermion point. In particular, we exploit a set of $T-Q$ equations of an inhomogeneous open spin-$\frac{1}{2}$ XXZ spin chain to compute the Casimir energy. We utilize a set of
auxiliary functions derived from these $T-Q$ equations to investigate the ground state of the bsG model at the free-Fermion point. We further compute the effective central charge 
in the ultraviolet (UV) limit. Our motivation here is three-fold: 
We first note that a detail study has been made in \cite{AN} on the free-Fermion point of the bsG model utilizing the solution of an open spin-$\frac{1}{2}$ XXZ spin chain. 
This is done for the case where a certain (complex) constraint is obeyed by the lattice boundary parameters. To our knowledge, such a study of the bsG model at the 
free-Fermion point  made by exploiting the $T-Q$ equations of an open spin-$\frac{1}{2}$ XXZ chain where the boundary terms are not subjected to any such constraint(s) 
has not been considered before. In addition, the $T-Q$ equations used in this paper contain two $Q(u)$ functions. A conventional $T-Q$ equation typically contains just 
one such $Q(u)$ function. The presence of multiple $Q(u)$ functions therefore requires a more elaborate analysis. We thus feel that the analysis presented in 
this paper which results from having two $Q(u)$ functions may serve as a guide and shall prove useful in studying the free-Fermion point of the bsG model with 
more general boundary terms. This requires a solution of the open spin-$\frac{1}{2}$ XXZ spin chain such as that given in \cite{MNC}.
Moreover, the presence of two $Q(u)$ functions for the more general case of $\mu = \pi/p$, where 
$p = 4\,,6\,,8\,,\ldots$\footnote{See \cite{Mureven1}} should lead to a set of nonlinear integral equations, crucial in determining the ground state 
Casimir energy of the bsG model. While the analysis can be relatively more involved, we feel that the present work at the free-Fermion point would better elucidate 
the process required in analyzing the case where $\mu = \pi/4\,,\pi/6\,,\pi/8\,,\dots$. 

This paper is arranged as follows: In Section 2, we review the bsG model. An open spin-$\frac{1}{2}$ XXZ chain with nondiagonal boundary terms and its $T-Q$ equations are reviewed next.  
In Section 3, the $T-Q$ equations of the corresponding inhomogeneous open spin-$\frac{1}{2}$ XXZ chain are reviewed.
This is followed by derivations of crucial auxiliary functions at the free-Fermion point. Utilizing the auxiliary functions 
and the analyticity of the transfer matrix eigenvalues, we present the derivation
of the Casimir energy at the free-Fermion point in Section 4 and compare our result for the Casimir energy of bsG model at the free-Fermion point with that 
of the TBA approach given in \cite{CSS2}. We then examine the result in the UV limit. The computation of the effective central charge in the UV limit is 
indeed a new result. We find it to be independent of the boundary parameters. We remark here that the approach using $T-Q$ equations can be viewed as an 
alternative way (to the elegant TBA approach employed in \cite{CSS2}) of arriving at the ground state Casimir energy. In \cite{CSS2}, the authors pointed out 
some of the technical difficulties encountered during the computations of the $K$ matrices which are indeed crucial in the ground state Casimir energy calculation. 
Computations of $K$ matrices there required complex analytic continuations. Nevertheless, this was carried out successfully in \cite{CSS2} to yield the desired results. 
Finally, plots of effective central charge for intermediate volume values are presented 
to support the analytical results obtained earlier for the UV limit. We conclude with a brief discussion of our results
and some open problems in Section 5.

\section{The bsG model and the $T-Q$ equations of an open spin-$\frac{1}{2}$ XXZ chain}\label{sec:ITQ}

In this section, we briefly review the bsG model (as given in \cite{AN}). 
The Euclidean action of the model on the finite ``spatial'' interval $x \in \left[ x_{-} \,, x_{+} \right]$ is described by 
\be
{\cal S} = \int_{-\infty}^{\infty}dy 
\int_{x_{-}}^{x_{+}}dx\  {\cal A}(\varphi \,, \partial_{\nu} \varphi) 
+ \int_{-\infty}^{\infty}dy \left[ 
{\cal B}_{-}(\varphi \,, {d\varphi\over dy} )\Big\vert_{x=x_{-}} +
{\cal B}_{+}(\varphi \,, {d\varphi\over dy} )\Big\vert_{x=x_{+}} \right] \,,
\label{SGaction}
\ee 
where the bulk action is given by 
\be 
{\cal A}(\varphi \,, \partial_{\nu} \varphi) = 
{1\over 2}(\partial_{\nu} \varphi)^{2}
+ \mu_{bulk} \cos (\beta \varphi) \,,
\label{SGbulkaction}
\ee 
and the boundary action is given by 
\be
{\cal B}_{\pm}(\varphi \,, {d\varphi\over dy} ) = 
\mu_{\pm} \cos( {\beta\over 2} (\varphi - \varphi_{0}^{\pm}))
\pm {\pi\gamma_{\pm}\over \beta} {d\varphi\over dy}  \,.
\label{SGboundaction}
\ee
As noted in \cite{AN, ABNPT}, the action is similar to the one considered by Ghoshal and Zamolodchikov \cite{GZ}, except for two boundaries
instead of one. In addition to that, the presence of an additional term depending on the ``time'' derivative of the field in the boundary action (\ref{SGboundaction}) is also 
well noted which as pointed out in \cite{AN}, can be eliminated in the one-boundary case by including a term proportional to 
$\partial_{x} \partial_{y} \varphi$ in (\ref{SGbulkaction}). This would result in the elimination of only one of the 
two $\gamma_{\pm}$ parameters (say, $\gamma_{+}$) and in a shift of the other ($\gamma_{-} \mapsto \gamma_{-} - \gamma_{+}$)
in the two-boundary case. The continuum bulk coupling constant $\beta$ is related to the lattice bulk parameter $\mu$ by $\beta^{2} = 8(\pi - \mu)$. 
The free-Fermion point thus corresponds to $\beta^{2} = 4\pi$ (or $\mu = \pi/2$). Further, the boundary parameters in the continuum 
action ($\mu_{\pm} \,, \varphi_{0}^{\pm}$) is related to the boundary parameters of the lattice model ($\alpha_{\pm} \,, \beta_{\pm}$) in the Hamiltonian of the open 
spin-$\frac{1}{2}$ XXZ quantum spin chain (see (\ref{Hamiltonian}) and footnote 2 below). In subsequent sections, we shall focus on the ground state 
of this model as a function of the interval length $L \equiv x_{+} - x_{-}$.  
In particular, we shall concentrate on the computation of the Casimir energy (order $1/L$) 
for the ground state. The length $L$ and the soliton mass $m$ are given by
\be
L = N \Delta \,, \qquad m={2\over \Delta} e^{-2\Lambda} \,,
\label{continuumlimit}
\ee
respectively. The relation of $m$ and $L$ to $\mu_{bulk}$ is known \cite{AZ2}. In (\ref{continuumlimit}), $\Delta$ is the lattice spacing, which in the continuum limit, taken to be $\Delta\rightarrow 0$ together with 
$\Lambda\rightarrow \infty$ and $N\rightarrow \infty$ for the inhomogeneity parameter and number of lattice sites respectively.

Next, we shall review the $T-Q$ equations of an open spin-$\frac{1}{2}$ XXZ chain with nondiagonal boundary terms \cite{Mureven1} 
whose Hamiltonian can be written as
\be
{\mathcal H} &=& {1\over 2}\sum_{n=1}^{N-1}\left( 
\sigma_{n}^{x}\sigma_{n+1}^{x}+\sigma_{n}^{y}\sigma_{n+1}^{y} 
+\cosh \eta\ \sigma_{n}^{z}\sigma_{n+1}^{z}\right) \non \\
&+& {1\over 2} \sinh \eta \left( 
\csch \alpha_{-} \sigma_{1}^{x} 
+ \csch \alpha_{+} \sigma_{N}^{x} \right) \,, \label{Hamiltonian} 
\ee
where $\sigma^{x}$, $\sigma^{y}$, $\sigma^{z}$ are the usual
Pauli matrices, $\eta$ is the bulk anisotropy parameter and $\alpha_{\pm}$ are the free boundary 
parameters. $\alpha_{\pm}$ are restricted to be pure imaginary to ensure the Hermiticity of the  Hamiltonian. $N$ is the number of spins/sites.\footnote{We note that this model does not represent the most general case of nondiagonal boundary terms. It contains only two free boundary parameters
instead of the usual six for the general case (see, e.g., \cite{MNC}). The parameters $\beta_{\pm}\,, \theta_{\pm}$ are taken to be zero.} Henceforth, in results that follow, 
we have used $\eta = i\pi/2$ (thus $\mu = \pi/2$ since $\eta = i\mu$), which corresponds to the free-Fermion point. The above Hamiltonian therefore reduces into,
\be
{\mathcal H} &=& {1\over 2}\sum_{n=1}^{N-1}\left( 
\sigma_{n}^{x}\sigma_{n+1}^{x}+\sigma_{n}^{y}\sigma_{n+1}^{y}\right) \non \\
&+& {i\over 2}  \left( 
\csch \alpha_{-} \sigma_{1}^{x} 
+ \csch \alpha_{+} \sigma_{N}^{x} \right) \,. \label{HamiltonianXX} 
\ee
The transfer-matrix eigenvalues $T(u)$ of the open spin-$\frac{1}{2}$ XXZ chain described by the Hamiltonian (\ref{HamiltonianXX}) satisfy a set of $T-Q$ equations given by 
\cite{Mureven1}
\be
T(u) &=& \left(\tilde{h}(u-i\frac{\pi}{4}) + \tilde{h}(u+i\frac{\pi}{4})\right)\frac{Q_{1}(u+i\frac{\pi}{2})}{Q_{2}(u)}\non\\
&=& \left(\tilde{h}(u-i\frac{\pi}{4})g(u-i\frac{\pi}{4}) + \tilde{h}(u+i\frac{\pi}{4})g(u+i\frac{\pi}{4})\right)\frac{Q_{2}(u+i\frac{\pi}{2})}{Q_{1}(u)}
\,, \label{TQsXX}
\ee
where
\be 
\tilde{h}(u) = \sinh^{2N+2}(u)\sinh(2u)\,,\non\\ 
g(u) = \cosh(u+\alpha_{-})\cosh(u-\alpha_{-})\cosh(u+\alpha_{+})\cosh(u-\alpha_{+}).\label{hgfunc}
\ee
We stress that (\ref{hgfunc}) holds only for even values of $N$. 
We note that there are two $Q(u)$ functions that make up the solution of this model, $Q_{1}(u)$ and $Q_{2}(u)$ whose zeros are the Bethe roots. These functions are given by
\be
Q_{1}(u) &=& \prod_{j=1}^{\frac{N}{2}+1} 
\sinh (u - u_{j}^{(1)}) \sinh (u + u_{j}^{(1)})\non\\
Q_{2}(u) &=& \prod_{j=1}^{\frac{N}{2}} 
\sinh (u - u_{j}^{(2)}) \sinh (u + u_{j}^{(2)})\,,
\label{Qbefstrhyp}
\ee
where $\{u_{k}^{(1)}, u_{k}^{(2)}\}$ are the Bethe roots (and the zeros of $Q_{1}(u)$ and $Q_{2}(u)$ respectively.)
These Bethe roots have the following structure for the ground state,
\be
\left\{ \begin{array}{ll}
v_{j}^{(a,1)}  & j = 1\,, 2\,, \ldots \,, \frac{N}{2}\\
{\pi\over 2} \lambda_{1}^{(1,2)} + {i \pi\over 2} & \\
\end{array} \right.\,, \qquad a = 1\,, 2 \,, 
\label{stringhypothesisIIc}
\ee
where $v_{j}^{(a,1)}$ and $\lambda_{1}^{(1,2)}$ are real. 
The $v_{j}^{(a,1)}$ are the zeros of $Q_{a}(u)$ that form real sea (``sea roots'') and ${\pi\over 2}\lambda_{1}^{(1,2)}$ is the 
real part of the single ``extra root'' (also a zero of $Q_{1}(u)$) which is not part of the ``seas''. We shall denote ${\pi\over 2}\lambda_{1}^{(1,2)}$ as simply $\lambda_{0}$
henceforth. Hence, there are two ``seas'' of real roots. The above structure for the Bethe roots is assumed to hold true in the $N\rightarrow\infty$ limit 
(the string hypothesis).\footnote{We urge the readers to refer to \cite{Mureven1} for details on this, including the Bethe ansatz equations which yield these roots as their solutions.}
Introducing (\ref{stringhypothesisIIc}) in (\ref{Qbefstrhyp}), we have the following form for the $Q_{a}(u)$ functions, where $a = 1\,,2$.
\be
Q_{1}(u) &=& \cosh (u - \lambda_{0})\cosh (u + \lambda_{0})\prod_{j=1}^{\frac{N}{2}} 
\sinh (u - v_{j}^{(1,1)}) \sinh (u + v_{j}^{(1,1)}) \,, \non \\
Q_{2}(u) &=& \prod_{j=1}^{\frac{N}{2}} 
\sinh (u - v_{j}^{(2,1)}) \sinh (u + v_{j}^{(2,1)})\,.
\label{QII}
\ee
Since we consider only the ground state, we shall use the form (\ref{QII}) for the $Q_{a}(u)$ functions in calculations that follow.  
We define here boundary parameters $a_{\pm}$ that are related to $\alpha_{\pm}$ (defined as $\alpha_{\pm} = \frac{i\pi}{2} a_{\pm}$). We stress here that, the Bethe roots for the ground state have the form
(\ref{stringhypothesisIIc}) only for the following values of the boundary parameters $a_{\pm}$ (see \cite{Mureven1}),
\be
{1\over 2} < |a_{\pm}| < {3\over 2} \,, 
\qquad a_{+} a_{-} > 0\,. 
\label{niceregionII}
\ee 

\section{Auxiliary functions at the free-Fermion point and the analyticity of transfer matrix eigenvalues}\label{sec:fF}

In this section, we shall review the $T-Q$ equations obeyed by the transfer
matrix eigenvalues of the corresponding inhomogeneous spin-$\frac{1}{2}$ XXZ quantum spin chain and give the auxiliary functions needed in the computation of the Casimir energy of  
the bsG model at the free-Fermion point. Method outlined in \cite{Su1, Su2} is used along with steps utilized in \cite{ANS}.  

\subsection{$T-Q$ equations of the inhomogeneous spin-$\frac{1}{2}$ XXZ spin chain and auxiliary functions}

We begin this section by presenting the $T-Q$ equations obeyed by the transfer matrix eigenvalues of the inhomogeneous spin-$\frac{1}{2}$ XXZ quantum spin chain 
(more accurately the XX chain, since $\eta = i\frac{\pi}{2}$. See (\ref{Hamiltonian}) and (\ref{HamiltonianXX}) in Sec. 2): 
\be
T(u) = \left(T_{A}^{(+)}(u) + T_{A}^{(-)}(u)\right)\frac{Q_{1}(u+i\frac{\pi}{2})}{Q_{2}(u)}
\,, \label{TQsXXinhom1}
\ee
and 
\be
T(u) = \left(T_{B}^{(+)}(u) + T_{B}^{(-)}(u)\right)\frac{Q_{2}(u+i\frac{\pi}{2})}{Q_{1}(u)}
\,, \label{TQsXXinhom2}
\ee
where 
\be
T_{A}^{(\pm)}(u) = \phi (u\pm i\frac{\pi}{4})\sinh (2u\pm i\frac{\pi}{2})\sinh^{2}(u\pm i\frac{\pi}{4})\,,\non\\
T_{B}^{(\pm)}(u) = \phi (u\pm i\frac{\pi}{4})\sinh (2u\pm i\frac{\pi}{2})\sinh^{2}(u\pm i\frac{\pi}{4})g^{(\pm)}(u)\,.
\label{TAB}
\ee
In (\ref{TAB}), $\phi(u) = \sinh^N(u-\Lambda)\sinh^N(u+\Lambda)$ and $g^{(\pm)}(u) = g(u\pm i\frac{\pi}{4})$ where $g(u)$ is given by (\ref{hgfunc}). 
$\Lambda$ is the inhomogeneity parameter. We note here that the functions $g^{(\pm)}(u)$ can be conveniently expressed in terms of 
$a_{\pm}$ by
\be
g^{(\pm)}(u) &=& \cosh(u+i\frac{\pi}{4}(2a_{-}\pm 1))\cosh(u-i\frac{\pi}{4}(2a_{-}\mp 1))\non\\
&\times& \cosh(u+i\frac{\pi}{4}(2a_{+}\pm 1))\cosh(u-i\frac{\pi}{4}(2a_{+}\mp 1)).
\label{gpm}
\ee 

Next, we define the following crucial auxiliary functions,
\be
a(u) &=& \frac{\sinh(2u+i\frac{\pi}{2})\phi(u+i\frac{\pi}{4})\sinh^{2}(u+i\frac{\pi}{4})}{\sinh(2u-i\frac{\pi}{2})\phi(u-i\frac{\pi}{4})\sinh^{2}(u-i\frac{\pi}{4})}\,,\non\\
b(u) &=& \frac{\sinh(2u+i\frac{\pi}{2})\phi(u+i\frac{\pi}{4})\sinh^{2}(u+i\frac{\pi}{4})g^{(+)}(u)}{\sinh(2u-i\frac{\pi}{2})\phi(u-i\frac{\pi}{4})\sinh^{2}(u-i\frac{\pi}{4})g^{(-)}(u)}\,. 
\label{auxfuncs}
\ee
As will be evident in the next section, these functions play an important role in the computation of the Casimir energy. Note that for real $u$, 
\be
\bar{a}(u) = a(-u) = \frac{1}{a(u)}\,,\qquad \bar{b}(u) = b(-u) = \frac{1}{b(u)}\,,\qquad g^{(+)}(-u) = g^{(-)}(u) = \bar{g}^{(+)}(u)\,.  
\label{auxfuncsprops}
\ee
In terms of these auxiliary functions, (\ref{TQsXXinhom1}) and (\ref{TQsXXinhom2}) can be rewritten in following forms,
\be
T(u) &=& \sinh(2u-i\frac{\pi}{2})\phi(u-i\frac{\pi}{4})\sinh^{2}(u-i\frac{\pi}{4})\frac{Q_{1}(u+i\frac{\pi}{2})}{Q_{2}(u)}(1 + a(u))\non\\
&=& \sinh(2u+i\frac{\pi}{2})\phi(u+i\frac{\pi}{4})\sinh^{2}(u+i\frac{\pi}{4})\frac{Q_{1}(u+i\frac{\pi}{2})}{Q_{2}(u)}(1 + \bar{a}(u))
\label{TQsXXinhom1b}
\ee
and
\be
T(u) &=& \sinh(2u-i\frac{\pi}{2})\phi(u-i\frac{\pi}{4})\sinh^{2}(u-i\frac{\pi}{4})g^{(-)}(u)\frac{Q_{2}(u+i\frac{\pi}{2})}{Q_{1}(u)}(1 + b(u))\non\\
&=& \sinh(2u+i\frac{\pi}{2})\phi(u+i\frac{\pi}{4})\sinh^{2}(u+i\frac{\pi}{4})g^{(+)}(u)\frac{Q_{2}(u+i\frac{\pi}{2})}{Q_{1}(u)}(1 + \bar{b}(u))
\label{TQsXXinhom2b}
\ee
respectively, from which the following Bethe Ansatz equations are evident,
\be
a(u_{k}^{(2)}) &=& -1 \non\\
b(u_{k}^{(1)}) &=& -1\,.
\ee

\subsection{Analyticity of the transfer matrix eigenvalues}

From the definitions of the auxiliary functions given in (\ref{auxfuncs}), it is clear that the point $u = 0$ is a zero of the transfer matrix eigenvalues 
defined in (\ref{TQsXXinhom1b}) and (\ref{TQsXXinhom2b}). We point out that $T(u)$ does not have zeros near the real axis except for this simple
zero at the origin. Thus, for the reason that will become evident later, one can remove the root of $T(u)$ at the origin by defining \cite{ANS},
\beq
\check{T}(u)={T(u)\over \mu(u)} \,, 
\label{analyT}
\eeq
where $\mu(u)$ is any function whose only real root is a simple zero at the origin,
that is $\mu(0)=0\,, \ \mu'(0)\ne 0$, where the prime denotes differentiation with respect to $u$. Consequently, the $T-Q$ equations (\ref{TQsXXinhom1b}) and (\ref{TQsXXinhom2b})
become
\be
\check{T}(u) = t_{-}^{A}(u)\, \frac{Q_{1}(u+i\frac{\pi}{2})}{Q_{2}(u)}A(u) = 
t_{+}^{A}(u)\, \frac{Q_{1}(u+i\frac{\pi}{2})}{Q_{2}(u)} \bar A(u)
\label{modifiedTQA}
\ee
and
\be
\check{T}(u) = t_{-}^{B}(u)\, \frac{Q_{2}(u+i\frac{\pi}{2})}{Q_{1}(u)}B(u) = 
t_{+}^{B}(u)\, \frac{Q_{2}(u+i\frac{\pi}{2})}{Q_{1}(u)} \bar B(u)
\label{modifiedTQB}
\ee
respectively, where
\be
t_{\pm}^{A}(u)={\sinh(2u \pm i\frac{\pi}{2} )\over \mu(u)}\,\sinh^{2}(u\pm i\frac{\pi}{4}) 
\phi(u \pm i\frac{\pi}{4}) \,\non\\
t_{\pm}^{B}(u)={\sinh(2u \pm i\frac{\pi}{2} )\over \mu(u)}\,\sinh^{2}(u\pm i\frac{\pi}{4}) 
\phi(u \pm i\frac{\pi}{4})g^{(\pm)}(u)\,.
\label{tpmAB}
\ee 
In (\ref{modifiedTQA}) and (\ref{modifiedTQB}), following definitions have been adopted, 
\be
A(u) = 1 + a(u)\,,\qquad \bar{A}(u) = 1 + \bar{a}(u)\non \\
B(u) = 1 + b(u)\,,\qquad \bar{B}(u) = 1 + \bar{b}(u)\,.
\label{defab}
\ee
Recalling (\ref{analyT}), we note that $\ln \check{T}(u)$ is analytic near the real axis. Thus, we have the following from Cauchy's theorem,
\beq
0=\oint_C du\ [\ln\check{T}(u)]'' e^{iku} \,,
\label{Cauchy}
\eeq
where the contour $C$ is chosen as in the Figure below, $\epsilon$ being small and positive.

\setlength{\unitlength}{0.8cm}
\begin{picture}(6,4)(-8,-2)
\put(-2.5,0){\line(1,0){5}}
\put(0,-1.5){\line(0,1){3}}
\put(-1.0,0.3){\vector(-1,0){0.2}}
\put(-2.0,0.3){\line(1,0){4}}
\put(1.0,-0.3){\vector(1,0){0.2}}
\put(-2.0,-0.3){\line(1,0){4}}
\put(0.75,0.5){$C_{1}$}
\put(0.75,-0.8){$C_{2}$}
\put(2.0,-0.3){\line(0,1){0.6}}
\put(-2.0,-0.3){\line(0,1){0.6}}
\put(2.2,0.1){$i\epsilon$}
\put(2.2,-0.3){$-i\epsilon$}
\put(-2.0,-2.2){$\textrm{Integration\, contour}$}
\end{picture}

\noindent Using (\ref{modifiedTQA}) and the first equation in (\ref{tpmAB}), (\ref{Cauchy}) can consequently be written as
\be
0 &=& \int_{C_1} du\ \left[ \ln t_{+}^{A}(u) \right]'' e^{iku}+
\int_{C_1} du\ \left\{ \ln \left[{Q_{1}(u+i\frac{\pi}{2})\over Q_{2}(u)}\right] \right\}'' e^{iku}
+\int_{C_1} du\ \left[ \ln \bar A(u) \right]'' e^{iku} \non\\
&+& \int_{C_2} du\ \left[ \ln t_{-}^{A}(u) \right]'' e^{iku} 
+ \int_{C_2} du\ \left\{ \ln\left[{Q_{1}(u+i\frac{\pi}{2})\over Q_{2}(u)}\right] 
\right\}'' e^{iku}
+\int_{C_2} du\ \left[ \ln A(u)\right]'' e^{iku} \,. \non\\ 
\label{Cauchy2TQA}
\ee
Following \cite{ANS}, we define Fourier transforms along $C_{2}$ and $C_{1}$ as 
\be
\widehat{Lf''}(k)=\int_{C_2} du\ [\ln f(u)]'' e^{iku} \,, \qquad
\widehat{{\cal L}f''}(k)=\int_{C_1} du\ [\ln f(u)]'' e^{iku} \,,
\label{fouriertransfdef}
\ee
respectively. Exploiting the periodicity \footnote{This is to make the imaginary part of the argument negative.}
\beq
Q_{a}(u)=Q_{a}(u-i\pi),\quad u\in C_1,\qquad{\rm and}\qquad
Q_{a}(u+i\frac{\pi}{2})=Q_{a}(u+i\frac{\pi}{2}-i\pi)=Q_{a}(u-i\frac{\pi}{2}),\quad u\in C_2 \,,
\label{Qperiodicity}
\eeq
where $a = 1\,,2$, and using the definitions (\ref{fouriertransfdef}), we arrive at the following forms for the second and fifth terms of (\ref{Cauchy2TQA}), 
\be
\int_{C_1}du\ \left\{ \ln \left[{Q_{1}(u+i\frac{\pi}{2})\over Q_{2}(u)}\right]
\right\}''e^{iku}&=&
-\widehat{LQ_{1s}''}(k)e^{-\frac{\pi}{2} k}+\widehat{LQ_{2}''}(k)e^{-\pi k}-2\pi k\frac{\cos(k\lambda_{0})}{\sinh(\frac{\pi k}{2})}e^{\frac{\pi k}{2}} \,, \non \\
\int_{C_2}du\ \left\{ \ln \left[{Q_{1}(u+i\frac{\pi}{2})\over Q_{2}(u)}\right] 
\right\}'' e^{iku}&=&
\widehat{LQ_{1s}''}(k)e^{-\frac{\pi}{2} k}-\widehat{LQ_{2}''}(k)+2\pi k\frac{\cos(k\lambda_{0})}{\sinh(\frac{\pi k}{2})}e^{\frac{\pi k}{2}} \,.
\label{FTQ12}
\ee
When calculating the Fourier transforms of $\left[\ln Q_{1}(u+i\frac{\pi}{2})\right]''$, we have separated the ``extra root'' contributions, thus resulting in
$\lambda_{0}$ dependent terms in (\ref{FTQ12}), where we recall that $\lambda_{0} + i\frac{\pi}{2}$ is a zero of $Q_{1}(u)$ (see (\ref{QII})). In (\ref{FTQ12}), $\widehat{LQ_{1s}''}(k)$ is defined as
\be
\widehat{LQ_{1s}''}(k)=\int_{C_2} du\ [\ln Q_{1s}(u)]'' e^{iku} \,, \qquad
\label{Q1sFT}
\ee
where
\be
Q_{1s}(u) &=& \prod_{j=1}^{\frac{N}{2}}\sinh (u - v_{j}^{(1,1)}) \sinh (u + v_{j}^{(1,1)})\,,
\label{Q1s}
\ee
which consists of just the ``sea roots''. Perfect cancellation of terms involving $\widehat{LQ_{1s}''}(k)$ and the ``extra root'' $\lambda_{0}$ is to be noted here. 
Next, defining
\be
C(k) \equiv \int_{C_1}du \left[ \ln t_{+}^{A}(u) \right]'' e^{iku}
+\int_{C_2}du \left[ \ln t_{-}^{A}(u) \right]'' e^{iku} \,,
\label{Ck}
\ee
and using (\ref{FTQ12}) (along with (\ref{Ck})) in (\ref{Cauchy2TQA}), we obtain the following,
\be
C(k) + \widehat{LQ_{2}''}(k)\left[e^{-\pi k}-1\right] + \widehat{LA''}(k)+\widehat{{\cal L}{\bar A}''}(k) = 0\,,
\ee
which in turn leads to the following result, 
\beq
\widehat{LQ_{2}''}(k)={e^{ \frac{\pi k}{2}}\over 2 \sinh( \frac{\pi k}{2})
}
\left[
\widehat{LA''}(k)+\widehat{{\cal L}{\bar A}''}(k)+C(k)\right] \,.
\label{logQ2}
\eeq
Repeating the above steps, now using (\ref{modifiedTQB}) and the second equation in (\ref{tpmAB}), we arrive at the following result,
\beq
\widehat{LQ_{1s}''}(k)={e^{ \frac{\pi k}{2}}\over 2 \sinh( \frac{\pi k}{2})
}
\left[
\widehat{LB''}(k)+\widehat{{\cal L}{\bar B}''}(k)+C_{B}(k)\right] \,,
\label{logQ1s}
\eeq
where 
\be
C_{B}(k) \equiv \int_{C_1}du \left[ \ln t_{+}^{B}(u) \right]'' e^{iku}
+\int_{C_2}du \left[ \ln t_{-}^{B}(u) \right]'' e^{iku} \,.
\label{CkB}
\ee
Results given in (\ref{logQ2}) and (\ref{logQ1s}) constitute the main consequence of analyticity and are significant in the calculation of the Casimir energy in the 
next section.

Given the appearance of terms $\widehat{LA''}(k)$, $\widehat{{\cal L}{\bar A}''}(k)$, $\widehat{LB''}(k)$ and $\widehat{{\cal L}{\bar B}''}(k)$ in (\ref{logQ2}) and 
(\ref{logQ1s}), it is natural to proceed with the calculations of $\widehat{La''}(k)$ and $\widehat{Lb''}(k)$ as the next step. We shall start with the former.
Taking the Fourier transform of $\left[\ln a(u)\right]''$ (utilizing (\ref{auxfuncs})) along $C_{2}$, one arrives at
\be
\widehat{La''}(k)&=&\int_{C_2}du\ [\ln a(u)]'' e^{iku}\non\\
&=&\int_{C_2}du\ \left\{
\ln\left[{\sinh(2u+i\frac{\pi}{2})\, \phi(u+i\frac{\pi}{4})\, \sinh^{2}(u+i\frac{\pi}{4})\over
\sinh(2u-i\frac{\pi}{2})\, \phi(u-i\frac{\pi}{4})\, \sinh^{2}(u-i\frac{\pi}{4})}\right] 
\right\}'' e^{iku}\,,
\ee
which upon evaluation yields,
\beq
\widehat{La''}(k)= -2\pi k \Bigg\{ {N\cos(\La k)\over \cosh({\pi k\over 4})} + \frac{1}{\cosh({\pi k\over 4})}\Bigg\}\,.
\label{afuncFouriersp}
\eeq
In obtaining the above, we have made use of the following identities (see \cite{ANS}),
\beq
\int_{C_2} {du\over 2\pi}  \left[ \ln \sinh(u-i \alpha)\right]'' e^{iku}= 
e^{-k(\alpha - n \pi)}  \psi(k) \,, 
\label{psi}
\eeq
where $n$ is an integer such that $0 < \Re e(\alpha - n \pi) < \pi$, 
and
\beq
\int_{C_2} {du\over 2\pi} \left[ \ln \sinh(2u)\right]'' e^{iku}
=\psi_2(k) \,,
\label{psi2}
\eeq
where $\psi(k) \equiv {k\over 1-e^{-\pi k}}$ and  $\psi_2(k) \equiv {k\over 1-e^{-{\pi k\over 2}}}$. Converting (\ref{afuncFouriersp}) to coordinate space and 
integrating twice yields the following, 
\be
\ln a(u) = -i 2N \tan^{-1}\left({\sinh (2u)\over 
\cosh (2\La) } \right) + i\, P_{bdry}(u) +C i\pi \,,
\label{afunccoordspace}
\ee
where $P_{bdry}(u)$ is given by
\be
P_{bdry}(u) = \int_{0}^{u}du'\, R(u') = \frac{1}{2} 
\int_{-u}^{u}du'\, R(u')\,,
\label{spinhalfPbdry}
\ee
and $R(u)$ refers to the Fourier transform of $\hat R(k)$ which is given below,
\be
\hat R(k) &=& -\frac{2\pi}{\cosh{\pi k\over 4}}.
\label{RR}
\ee
The integration constant $C$ in (\ref{afunccoordspace}) is obtained by considering the $u\rightarrow\infty$ limit of (\ref{afunccoordspace}) and the function $a(u)$
as given by (\ref{auxfuncs}). Proceeding as in \cite{ANS}, one obtains the correct factor, $C=1$. 
Taking the continuum limit ($\La \rightarrow \infty\,, N\rightarrow\infty\,, \Delta\rightarrow 0)$, the term 
$-i2N \tan^{-1}\left({\sinh (2u)\over 
\cosh (2\La)} \right)$ becomes $-i 2mL \sinh \theta$ after defining the renormalized rapidity $\theta$ as
\be 
\theta = 2u \,.
\label{renormrapidity}
\ee 
(\ref{afunccoordspace}) therefore becomes
\be
\ln \af(\theta) = -i 2mL \sinh \theta + i\, \Pf_{bdry}(\theta) +i\pi\,,
\label{afunctheta}
\ee
where following definitions have been used,
\be
\af(\theta)=a(\frac{\theta}{2})\,, \quad \Pf_{bdry}(\theta)=P_{bdry}(\frac{\theta}{2})\,.  
\label{mathnewaPdefs}
\ee 
The above maneuver can now be repeated for $\widehat{Lb''}(k)$, using $b(u)$ defined in (\ref{auxfuncs}).
The Fourier transform of $\left[\ln b(u)\right]''$ along $C_{2}$ gives,
\be
\widehat{Lb''}(k)&=&\int_{C_2}du\ [\ln b(u)]'' e^{iku}\non\\
&=&\int_{C_2}du\ \left\{
\ln\left[{\sinh(2u+i\frac{\pi}{2})\, \phi(u+i\frac{\pi}{4})\, \sinh^{2}(u+i\frac{\pi}{4})\, g^{(+)}(u)\over
\sinh(2u-i\frac{\pi}{2})\, \phi(u-i\frac{\pi}{4})\, \sinh^{2}(u-i\frac{\pi}{4})\, g^{(-)}(u)}\right] 
\right\}'' e^{iku}\,,
\ee
which upon evaluation as before yields,
\be
\widehat{Lb''}(k) &=& -2\pi k \Bigg\{ {N\cos(\La k)\over \cosh({\pi k\over 4})} + \frac{1}{\cosh({\pi k\over 4})} \non \\
&-& \frac{1}{\sinh({\pi k\over 2})}
\left[\sinh ({\pi k\over 4}B_{-}) + \sinh ({\pi k\over 4}B_{+}) - \sinh ({\pi k\over 4}A_{-}) - \sinh ({\pi k\over 4}A_{+})\right]\Bigg\}\,, \non\\
\label{bfuncFouriersp}
\ee
where identities (\ref{psi}) and (\ref{psi2}) have been utilized. In (\ref{bfuncFouriersp}), $A_{\pm} = 2a_{\pm}- 1$ and $B_{\pm} = 2a_{\pm} + 1$.
Converting (\ref{bfuncFouriersp}) to coordinate space and integrating twice results in the following, 
\be
\ln b(u) = -i 2N \tan^{-1}\left({\sinh (2u)\over 
\cosh (2\La) } \right) + i\, P_{bdry}^{B}(u) + D i\pi \,,
\label{bfunccoordspace}
\ee
where $P_{bdry}^{B}(u)$ is given by
\be
P_{bdry}^{B}(u) = \int_{0}^{u}du'\, R_{B}(u') = \frac{1}{2} 
\int_{-u}^{u}du'\, R_{B}(u')\,,
\label{spinhalfPbdryB}
\ee
and $R_{B}(u)$ refers to the Fourier transform of $\hat R_{B}(k)$ which is given below,
\be
\hat R_{B}(k) &=& -\frac{2\pi}{\cosh{\pi k\over 4}} + \frac{2\pi}{\sinh({\pi k\over 2})}\left[\sinh ({\pi k\over 4}B_{-}) + \sinh ({\pi k\over 4}B_{+}) - \sinh ({\pi k\over 4}A_{-})
- \sinh ({\pi k\over 4}A_{+})\right]\,.\non\\
\label{Rb}
\ee
The integration constant $D = -1$ in (\ref{bfunccoordspace}) is obtained in a similar manner as for $C$ in (\ref{afunccoordspace}). 
Taking the continuum limit ($\La \rightarrow \infty\,, N\rightarrow\infty\,, \Delta\rightarrow 0)$, gives the following result for $\bff(\theta)$  
in terms of the renormalized rapidity $\theta$,
\be
\ln \bff(\theta) = -i 2mL \sinh \theta + i\, \Pf_{bdry}^{B}(\theta) - i\pi\,,
\label{bfunctheta}
\ee
where following definitions are adopted,
\be
\bff(\theta)=b(\frac{\theta}{2})\,, \quad \Pf_{bdry}^{B}(\theta)=P_{bdry}^{B}(\frac{\theta}{2})\,.  
\label{mathnewbPdefs}
\ee 

\section{Casimir energy}\label{sec:Casimir}

In this section, we give the main result of this paper. We compute the Casimir energy (order $1/L$) for the ground state of a bsG model. Following the prescription of 
Reshetikhin and Saleur \cite{RS} (see also \cite{ANS}), the energy for the inhomogeneous case 
($\Lambda \ne 0$) of a spin-$\frac{1}{2}$ XXZ quantum spin chain is given by 
\beq
E=-\frac{g}{\Delta}\left\{ \frac{d}{du}\ln T(u)\Bigg\vert_{u=\Lambda+\frac{i\pi}{4}}
-\frac{d}{du}\ln T(u)\Bigg\vert_{u=\Lambda-\frac{i\pi}{4}}\right\} 
\,, \label{energydef}
\eeq
where $g$ is given by 
\be
g=-\frac{i}{8} \,. \label{energynormalization}
\ee 
Using the fact that
\be 
\frac{d}{du}\ln T(u)\Bigg\vert_{u=\Lambda\pm\frac{i\pi}{4}}=
\frac{d}{du}\ln 
T^{(\pm)}(u)\Bigg\vert_{u=\Lambda\pm\frac{i\pi}{4}} \,
\label{deff2}
\ee
where
\be
T^{(\pm)}(u)&=& T_{A}^{(\pm)}(u)\frac{Q_{1}(u + \frac{i\pi}{2})}{Q_{2}(u)}\,,
\label{TpmA}
\ee
and $T_{A}^{(\pm)}(u)$ is given by (\ref{TAB}), one can recast (\ref{energydef}) as follows,
\be
E&=&-\frac{g}{\Delta}\frac{d}{du}\left\{ \ln T^{(+)}(u+\frac{i\pi}{4})-
\ln T^{(-)}(u-\frac{i\pi}{4})\right\} \bigg\vert_{u=\Lambda}\,,
\label{energydef2}
\ee
which in turn yields,
\be
E&=& -\frac{g}{\Delta}\int \frac{dk}{2\pi} e^{-ik\Lambda}\left[
e^{\frac{\pi k}{4}}\widehat{LT^{(+)'}}(k)-e^{-\frac{\pi k}{4}}\widehat{LT^{(-)'}}(k)\right] \,,
\label{FTenergy}
\ee
after utilizing the following definition,
\be
[\ln f(u)]' = \int \frac{dk}{2\pi}\ \widehat{Lf'}(k)\ e^{-ik u} \,, \qquad
u \in C_{2} \,,
\label{FTfact}
\ee 
that follows from (\ref{fouriertransfdef})).
 
The calculation of the quantity in the square bracket of the integrand of (\ref{FTenergy}) is now in order. Using the result
\beq
\widehat{LT^{(\pm)'}}(k)=\int_{C_2} du\ [\ln T^{(\pm)}(u)]' e^{iku} 
\label{fouriertransfdeffprime}
\eeq
together with (\ref{TAB}), (\ref{psi}), (\ref{psi2}) and (\ref{TpmA}), one finds the following
\be
e^{\frac{\pi k}{4}}\widehat{LT^{(+)'}}(k)&-&e^{-\frac{\pi k}{4}}\widehat{LT^{(-)'}}(k) \non \\
&=&
(1 - e^{-\frac{\pi k}{2}}) \frac{2\pi\psi_2(k)}{(-ik)} + 2e^{-\frac{\pi k}{2}}\sinh(\frac{\pi k}{4})\left[\widehat{LQ_{1s}'}(k) + \widehat{LQ_{2}'}(k)\right]\non \\
&+& 2\pi i\frac{\cos(k\lambda_{0})}{\cosh(\frac{\pi k}{4})}e^{\frac{\pi k}{2}} \,.
\label{tpmdiffA}
\ee
Repeating the above steps using 
\be
T^{(\pm)}(u)&=& T_{B}^{(\pm)}(u)\frac{Q_{2}(u + \frac{i\pi}{2})}{Q_{1}(u)}\,,
\label{TpmB}
\ee
instead yields (Note that $T_{B}^{(\pm)}(u)$ is also given by (\ref{TAB}).),
\be
e^{\frac{\pi k}{4}}\widehat{LT^{(+)'}}(k)&-&e^{-\frac{\pi k}{4}}\widehat{LT^{(-)'}}(k) \non \\
&=&
(1 - e^{-\frac{\pi k}{2}}) \frac{2\pi\psi_2(k)}{(-ik)} + 2e^{-\frac{\pi k}{2}}\sinh(\frac{\pi k}{4})\left[\widehat{LQ_{1s}'}(k) + \widehat{LQ_{2}'}(k)\right]\non \\
&+& e^{\frac{\pi k}{4}}\widehat{Lg^{(+)'}}(k) - e^{-\frac{\pi k}{4}}\widehat{Lg^{(-)'}}(k) - 2\pi i\frac{\cos(k\lambda_{0})}{\cosh(\frac{\pi k}{4})} \,,
\label{tpmdiffB}
\ee
where 
\beq
\widehat{Lg^{(\pm)'}}(k)=\int_{C_2} du\ [\ln g^{(\pm)}(u)]' e^{iku} 
\label{fouriertransfdeffprime}\,.
\eeq
We recall that $g^{(\pm)}(u)$ are defined in (\ref{gpm}). Prior to determining the Casimir energy from (\ref{FTenergy}), we first need to determine the explicit 
form for $\widehat{LQ_{1s}'}(k) + \widehat{LQ_{2}'}(k)$. Next, the ``extra root'' terms in (\ref{tpmdiffA}) and (\ref{tpmdiffB}) need to be addressed. Fortunately,
the latter can be managed by simply solving for $2\pi i\frac{\cos(k\lambda_{0})}{\cosh(\frac{\pi k}{4})}$ by subtracting (\ref{tpmdiffA}) from (\ref{tpmdiffB}). This
yields the following,
\be
2\pi i\frac{\cos(k\lambda_{0})}{\cosh(\frac{\pi k}{4})} = \frac{e^{\frac{\pi k}{4}}\widehat{Lg^{(+)'}}(k) - e^{-\frac{\pi k}{4}}\widehat{Lg^{(-)'}}(k)}{1 + e^{\frac{\pi k}{2}}}\,.
\label{extrarootintpmdiff}
\ee 
To determine the sum $\widehat{LQ_{1s}'}(k) + \widehat{LQ_{2}'}(k)$, we recall (\ref{logQ2}), (\ref{logQ1s}) and use 
$\widehat{LQ_{l}'}(k) = {1\over (-i k)}\widehat{LQ_{l}''}(k)$ where $l = 1s\,,2$. Thus, we obtain
\be
\widehat{LQ_{1s}'}(k) + \widehat{LQ_{2}'}(k) = -\frac{e^{ \frac{\pi k}{2}}}{2i k \sinh( \frac{\pi k}{2})}
\Big\{
\widehat{LA'}(k)+\widehat{{\cal L}{\bar 
A}'}(k)+\widehat{LB'}(k)+\widehat{{\cal L}{\bar 
B}'}(k) + C(k) +C_{B}(k)\Big\}\,,\non \\
\label{LQsum1}
\ee
where $C(k)$ and $C_{B}(k)$ are given by (\ref{Ck}) and (\ref{CkB}) respectively, which are evaluated to be,
\be
C(k) &=& 2\pi k \Bigg\{ {N\cos(\La k)\over \cosh({\pi k\over 4})} + \frac{1}{\cosh({\pi k\over 4})} - 1\Bigg\}\non \\
C_{B}(k) &=& 2\pi k \Bigg\{ {N\cos(\La k)\over \cosh({\pi k\over 4})} + \frac{1}{\cosh({\pi k\over 4})} -1 \non \\
&-& \frac{1}{\sinh({\pi k\over 2})}
\left[\sinh ({\pi k\over 4}B_{-}) + \sinh ({\pi k\over 4}B_{+}) - \sinh ({\pi k\over 4}A_{-}) - \sinh ({\pi k\over 4}A_{+})\right]\Bigg\}\,.\non\\
\label{CandCB}
\ee
In evaluating $C(k)$ and $C_{B}(k)$, the following result has been used, 
\be
\oint_{C}du\
\left[\ln\mu(u)\right]''e^{iku} = 2\pi k\,.
\label{intofmu}
\ee
Subsequently, substituting (\ref{extrarootintpmdiff})-(\ref{CandCB}) into (\ref{tpmdiffA}) or (\ref{tpmdiffB}) results in,
\be
e^{\frac{\pi k}{4}}\widehat{LT^{(+)'}}(k)&-&e^{-\frac{\pi k}{4}}\widehat{LT^{(-)'}}(k) \non \\
&=& \frac{1}{2\cosh(\frac{\pi k}{4})}\left[\widehat{LA'}(k)+\widehat{{\cal L}{\bar A}'}(k)+\widehat{LB'}(k)+\widehat{{\cal L}{\bar B}'}(k)\right]\non\\
&+& 2i\pi\left[1-\frac{1}{\cosh (\frac{\pi k}{4})} + \frac{1}{\cosh^{2}(\frac{\pi k}{4})}\right] + \tanh (\frac{\pi k}{4})e^{\frac{\pi k}{4}}\widehat{Lg^{(+)'}}(k) \non\\
&+& 2e^{-\frac{\pi k}{2}}\tanh (\frac{\pi k}{4})\widehat{L\phi'}(k)  \,,
\label{tpmdifffinal}
\ee 
which upon substitution in (\ref{FTenergy}) eventually leads to the following result for the Casimir energy for the ground state,\footnote{(\ref{FTenergy})
also contains the bulk and the boundary terms for the energy, which are not addressed here.}
\be
E_{C} = -\frac{g}{2\pi \Delta}\int_{-\infty}^{\infty} dk\ e^{-ik\Lambda}\frac{1}{2\cosh{\pi k\over{4}}}\left[
\widehat{LA'}(k)+\widehat{{\cal L}{\bar A}'}(k)+\widehat{LB'}(k)+\widehat{{\cal L}{\bar B}'}(k)\right]\,.
\label{Cas} 
\ee
In (\ref{tpmdifffinal}), $\widehat{L\phi'}(k) = \int_{C_2} du\ [\ln \phi (u)]' e^{iku}$.
We shall evaluate $E_{C}$ explicitly next.  
Passing to coordinate space and taking the continuum limit, (\ref{Cas}) becomes
\be
E_C&=&\frac{2g}{i\Delta\pi}
\int_{-\infty}^{\infty}du\,\Im m \left(\frac{1}{\cosh 2
(\Lambda-u-i\ep)}\right)'\left[\ln {\bar A}(u+i\ep)+\ln {\bar B}(u+i\ep)\right]\,.
\ee
Moreover, we have $\left(\frac{1}{\cosh 2(\Lambda-u-i\ep)}\right)'\rightarrow 4e^{-2(\Lambda-u-i\epsilon)}$ at 
$\Lambda\rightarrow\infty$ limit. Using (\ref{continuumlimit}), we have
\be
E_{C} = {1\over 2} \Big\{-\frac{m}{\pi}
\int_{-\infty}^{\infty}du\, \Im m \, e^{2(u+i\ep)}
\ln \bar A(u+i\ep) -\frac{m}{\pi}
\int_{-\infty}^{\infty}du\, \Im m \, e^{2(u+i\ep)}
\ln \bar B(u+i\ep)\Big\}\,.
\ee
Further, using (\ref{defab}), (\ref{renormrapidity}), (\ref{mathnewaPdefs}) and (\ref{mathnewbPdefs}) and after some manipulation, $E_{C}$ reduces to, 
\be
E_{C} = -\frac{m}{4\pi}
\int_{-\infty}^{\infty}d\theta\, \Im m \, \sinh(\theta+i\varepsilon)\left[\ln \bar \Af(\theta+i\varepsilon) + \ln \bar \Bf(\theta+i\varepsilon)\right]\,,
\label{Casimir2}
\ee
where  
\be
\varepsilon= 2\epsilon\,, \quad \!
\Af(\theta)=A(\frac{\theta}{2})\,, \quad \!
\Bf(\theta)=B(\frac{\theta}{2}) \quad \!\,.
\label{epsABdef}
\ee
(Note also that we have used ${\bar \Af}(u)=\Af(-u)$, ${\bar \Bf}(u)=\Bf(-u)$ and $\Im m z=-\Im m{\bar z}$ in the process.)
Finally, introducing the identifications 
\be
{\bar \Af}(\theta) = 1 - e^{f_{a}(\theta)}\,,\qquad {\bar \Bf}(\theta) = 1 - e^{f_{b}(\theta)}\,, 
\label{identif}
\ee
the above becomes
\be
E_{C} = -{m\over 4\pi} \int_{-\infty}^{\infty} d\theta\ 
\Im m\ \sinh (\theta+ i \varepsilon)  \left[\ln (1 - e^{f_{a}(\theta + i \varepsilon)}) + \ln (1 - e^{f_{b}(\theta + i \varepsilon)})\right]
\,.
\label{Casimirfinal}
\ee

\subsection{Comparison with TBA results}

Next, we compare the result (\ref{Casimirfinal}) with the TBA results given in \cite{CSS2}. From (\ref{afunctheta}), (\ref{bfunctheta}) and (\ref{identif}), one finds
the following for $f_{a}(\theta)$ and $f_{b}(\theta)$,
\be
f_{a}(\theta) &=& 2i m L \sinh \theta - i \Pf_{bdry}(\theta) \non\\
f_{b}(\theta) &=& 2i m L \sinh \theta - i \Pf_{bdry}^{B}(\theta)\,.
\label{fafb}
\ee
We first rewrite (\ref{Casimirfinal}) as
\be
E_{C} &=& -{m\over 8\pi i} \Big\{\int_{-\infty}^{\infty} d\theta\ 
\sinh \theta\ \left[ \ln (1 - e^{f_{a}(\theta + i \varepsilon)})
- \ln (1 - e^{-f_{a}(\theta - i \varepsilon)})\right] \non\\
&+& \int_{-\infty}^{\infty} d\theta\ 
\sinh \theta\ \left[ \ln (1 - e^{f_{b}(\theta + i \varepsilon)})
- \ln (1 - e^{-f_{b}(\theta - i \varepsilon)})\right]\Big\}\,.
\label{Casimirfinal2}
\ee
Performing a change in integration variables of $\theta' = \theta - {i\pi\over
2} + i \varepsilon$ in the first and third terms, and $\theta' = \theta +
{i\pi\over 2}- i \varepsilon$ in the second and fourth terms, the above becomes
\be
E_{C} &=& -{m\over 8\pi} \Big\{\int_{-\infty}^{\infty} d\theta'\ 
\left[\cosh (\theta' - i\varepsilon) \ln (1 - e^{f_{a}(\theta' + i\frac{\pi}{2})})
+ \cosh (\theta' + i\varepsilon)\ln (1 - e^{-f_{a}(\theta' - i\frac{\pi}{2})})\right] \non\\
&+& \int_{-\infty}^{\infty} d\theta'\ 
\left[ \cosh (\theta' - i\varepsilon) \ln (1 - e^{f_{b}(\theta' + i\frac{\pi}{2})})
+ \cosh (\theta' + i\varepsilon)\ln (1 - e^{-f_{b}(\theta' - i\frac{\pi}{2})})\right]\Big\}\,.
\label{Casimirfinal3}
\ee
Upon the assumption that the resulting contours can then be deformed to the real axis, and dropping
the primes, we obtain
\be
E_{C} &=& -{m\over 8\pi} \int_{-\infty}^{\infty} d\theta\ 
\cosh \theta\ \ln \left[(1 - e^{f_{a}(\theta + {i\pi\over 2})})(1 - e^{-f_{a}(\theta - {i\pi\over 2})})
\right] \non \\ \non \\
&-& {m\over 8\pi} \int_{-\infty}^{\infty} d\theta\ 
\cosh \theta\ \ln \left[(1 - e^{f_{b}(\theta + {i\pi\over 2})})(1 - e^{-f_{b}(\theta - {i\pi\over 2})})
\right]\,. \non \\
\label{SGCasimirfinal}
\ee 
Using (\ref{fafb}) along with equations (\ref{spinhalfPbdry}), (\ref{RR}), (\ref{spinhalfPbdryB}) and (\ref{Rb}), the following results are obtained,
\be
e^{-i\Pf_{bdry}(\theta + i\frac{\pi}{2})} &=& -\tanh^{2}(\frac{\theta}{2})\non\\
e^{-i\Pf_{bdry}^B(\theta + i\frac{\pi}{2})} &=& -\tanh^{2}(\frac{\theta}{2})\coth(\frac{\theta}{2} + i\frac{\pi}{2}a_{-})\coth(\frac{\theta}{2} - i\frac{\pi}{2}a_{-})\non\\
&\times& \coth(\frac{\theta}{2} + i\frac{\pi}{2}a_{+})\coth(\frac{\theta}{2} - i\frac{\pi}{2}a_{+})\,.
\label{epbdry}
\ee
Further, utilizing
\be
e^{-i\Pf_{bdry}(\theta + i\frac{\pi}{2})} &=& e^{i\Pf_{bdry}(\theta - i\frac{\pi}{2})}\non \\
e^{-i\Pf_{bdry}^{B}(\theta + i\frac{\pi}{2})} &=& e^{i\Pf_{bdry}^{B}(\theta - i\frac{\pi}{2})}\,,
\label{epbdryidentity}
\ee
we arrive at
\be
E_{C} =-{m\over 2\pi} \int_{0}^{\infty} d\theta\ 
\cosh \theta\  \ln \left( 1 + E_{1}(\theta)\ e^{-2m L \cosh \theta}
+ E_{2}(\theta)\ e^{-4m L \cosh \theta} \right) \,,
\label{SGCasimirFF}
\ee
where
\be
E_{1}(\theta) = -e^{-i \Pf_{bdry}(\theta + {i\pi\over 2})}
- e^{-i \Pf_{bdry}^{B}(\theta + {i\pi\over 2})} \,, \qquad
E_{2}(\theta) = e^{-i \Pf_{bdry}(\theta + {i\pi\over 2})}
e^{-i \Pf_{bdry}^{B}(\theta + {i\pi\over 2})} \,.
\label{e1e2}
\ee
We recall now the result obtained using the TBA approach in \cite{CSS2} (Eq. (58)), where the Casimir
energy is given by (\ref{SGCasimirFF}), with
\be
E_{1}(\theta) = \tr \left( {\bar K}_{-}(\theta)\ K_{+}(\theta) \right) 
\,, \qquad
E_{2}(\theta) = \det  \left( {\bar K}_{-}(\theta)\ K_{+}(\theta) \right) 
\,, \label{e1e2ofCSS}
\ee
where $K_{\pm}(\theta)$ are the boundary $S$ matrices
\cite{GZ}
\be
K_{\pm}(\theta) = r_{\pm}({i \pi\over 2} - \theta)
\left( \begin{array}{cc}
- {i k_{\pm}\over 2} e^{-i\gamma_{\pm}} \sinh 2\theta
& \sin (\xi_{\pm} - i\theta) \\
 -\sin (\xi_{\pm} + i\theta) 
& - {i k_{\pm}\over 2} e^{i\gamma_{\pm}} \sinh 2\theta
\end{array} \right) \,.
\label{boundSmatrix}
\ee
$r_{\pm}(\theta)$ refer to scalar factors, which are expounded below. The boundary $S$ matrices (as given by \cite{AN}), contain
the parameters $\gamma_{\pm}$ which correspond to the ${d\varphi\over dy}$ terms in the boundary action of the bsG model given in
(\ref{SGboundaction}). Moreover, the parameters $\gamma_{\pm}$ are the same as that appear in (\ref{SGboundaction}). We urge the readers to refer to footnote 8 in 
\cite{AN} for more details on this. We also remark here that $\gamma_{\pm}$ are conjectured in \cite{AN} to be related to the lattice parameters $\theta_{\pm}$, which
we have set to be zero in the Hamiltonians (\ref{Hamiltonian}) and (\ref{HamiltonianXX}). Therefore $\gamma_{\pm}$ here will be taken to be zero as well.
In addition, the Goshal-Zamolodchikov boundary parameters ($\xi_{\pm}, k_{\pm}$) is related to the continuum parameters $(\vartheta_{\pm},\eta_{\pm})$ by \cite{GZ}
\be
\cos\eta_{\pm} \cosh \vartheta_{\pm} = -\frac{1}{k_{\pm}}\cos\xi_{\pm}\,, \qquad \cos^{2}\eta_{\pm} + \cosh^{2}\vartheta_{\pm} = 1 + \frac{1}{k_{\pm}^{2}}\,.
\label{transform}
\ee
For the case studied here (recall footnote 2), the lattice parameters $\beta_{\pm} = 0$. Since the continuum parameters $\vartheta_{\pm}$ are related to $\beta_{\pm}$ 
through the equation $\vartheta_{\pm} = 2\beta_{\pm}$ at the free-Fermion point \cite{AN}, it follows that $\vartheta_{\pm} = 0$ here as well. Further, using (\ref{transform})
in (\ref{boundSmatrix}) and provided that
\be
(1-a_{\pm})\pi = \eta_{\pm}
\label{contlatticea}
\ee
which is also given in \cite{AN} for the case where contraints exist among these parameters (see Eqs. (2.8) and (3.22) of the reference \cite{AN}), one can then 
verify that (\ref{e1e2}) and (\ref{e1e2ofCSS}) agree for $E_{1}(\theta)$ and $E_{2}(\theta)$. Thus, this supports the notion that (\ref{contlatticea}) holds true even when the continuum paramaters 
$\eta_{\pm}$, are not subjected to constraints such as that given in the first line of Eq. (3.22) in \cite{AN}. Additionally, this also provides a check on the results 
(\ref{afunctheta}),(\ref{bfunctheta}) and (\ref{Casimir2}). Following conventions used in \cite{AN}, 
the scalar factors $r_{\pm}(\theta)$ in terms of the continuum parameters are given by
\be
r_{\pm}(\theta) = {1\over \cos \xi_{\pm}} 
\sigma(\eta_{\pm} \,, -i\theta)\ \sigma(i\vartheta_{\pm} \,, -i\theta)
\,,
\ee
where \cite{AKL}
\be
\sigma(x \,, u) = {\cos x\over 2 
\cos\left( {\pi\over 4} + {x\over 2} -{u\over 2} \right)
\cos\left( {\pi\over 4} - {x\over 2} -{u\over 2} \right)} \,.
\ee

\subsection{UV limit}

Now we consider the UV limit ($m L \rightarrow 0$) of the result (\ref{Casimirfinal}). First, using $E_{C} = -\frac{\pi c_{eff}}{24 L}$, where $c_{eff}$ is the 
effective central charge and (\ref{Casimirfinal}), we obtain the following for $c_{eff}$ with $\varepsilon \rightarrow 0$,
\be
c_{eff} = \frac{6mL}{\pi^{2}}\int_{-\infty}^{\infty} d\theta\ 
\Im m\ \sinh (\theta)  \left[\ln (1 - e^{f_{a}(\theta)}) + \ln (1 - e^{f_{b}(\theta)})\right]\,.
\label{ceffUV}
\ee
Next, realizing that only large values of $|\theta|$ contribute in the $m L \rightarrow 0$ limit, we first consider $\theta \gg 1$.
(\ref{ceffUV}) then becomes
\be
c_{eff,p} = \frac{3mL}{\pi^{2}}\int_{-\infty}^{\infty} d\theta\ 
\Im m\ e^{\theta}  \left[\ln (1 - e^{f_{a,p}(\theta)}) + \ln (1 - e^{f_{b,p}(\theta)})\right]\,.
\label{ceffUVpos}
\ee
The subscript $p$ is included to represent $c_{eff}\,, f_{a}(\theta)$ and $f_{b}(\theta)$ at $\theta \gg 1$. We follow closely steps in \cite{LMSS} by using 
results about dilogarithms (see \cite{KR}) defined by,
\be
L_{d}(x) \equiv \int_{0}^{x}dy\left[\frac{\ln (1+y)}{y} - \frac{\ln y}{1+y}\right]\,. 
\label{dilogs}
\ee
Thus, assuming\footnote{We note that the function $G$ in Eq.(7.34) of \cite{LMSS} is zero at the free-Fermion point, $\mu = \frac{\pi}{2}$.}
\be
-i \ln F(x) = \Psi(x)\,,
\label{Fpsi}
\ee
one has
\be
\int_{-\infty}^{\infty} dx 
\Im m\ \Psi'(x) \ln (1 + F(x + i\varepsilon)) &=&  \frac{1}{2} \Re e\ \Large(L_{d}[F(-\infty)] - L_{d}[F(\infty)]\Large)\non\\
&+& \frac{1}{2}\Im m\ \Large(\Psi(\infty) \ln (1 + F(\infty)) - \Psi(-\infty) \ln (1 + F(-\infty))\Large)\,. \non \\
\label{dilogsres}
\ee
Making the correspondence $F \leftrightarrow F_{l}\,,\Psi \leftrightarrow \Psi_{l,p}$ where $l = a\,,b$ and therefore $-i \ln F_{l}(\theta) = \Psi_{l,p}(\theta)$ 
with the following identifications
\be
F_{a}(\theta) = e^{f_{a,p}(\theta) - i\pi}\non \\
\Psi_{a,p}(\theta) = mLe^{\theta} - \Pf_{bdry}(\infty) - \pi\non\\
F_{b}(\theta) = e^{f_{b,p}(\theta) + i\pi}\non \\
\Psi_{b,p}(\theta) = mLe^{\theta} - \Pf_{bdry}^{B}(\infty) + \pi\,,\non\\
\label{CapFAPFBP}
\ee
where (\ref{fafb}) yields
\be
f_{a,p}(\theta) = imLe^{\theta} - i\Pf_{bdry}(\infty)\non\\
f_{b,p}(\theta) = imLe^{\theta} - i\Pf_{bdry}^{B}(\infty)\,,
\label{fapfbp}
\ee
one finds $c_{eff,p} = \frac{1}{2}$ from (\ref{ceffUVpos}). Similar computations for $\theta \ll 1$ yield $c_{eff, n} = \frac{1}{2}$. The subscript $n$ refers 
to $c_{eff}$ at $\theta \ll 1$. Similar results to (\ref{CapFAPFBP}) and (\ref{fapfbp}) have been used (for $\theta \ll 1$), namely
\be
F_{a}(\theta) = e^{f_{a,n}(\theta) - i\pi}\non \\
\Psi_{a,n}(\theta) = -mLe^{-\theta} - \Pf_{bdry}(-\infty) - \pi\non\\
F_{b}(\theta) = e^{f_{b,n}(\theta) + i\pi}\non \\
\Psi_{b,n}(\theta) = -mLe^{-\theta} - \Pf_{bdry}^{B}(-\infty) + \pi\,,\non\\
\label{CapFANFBN}
\ee
where we find
\be
f_{a,n}(\theta) = -imLe^{-\theta} - i\Pf_{bdry}(-\infty)\non\\
f_{b,n}(\theta) = -imLe^{-\theta} - i\Pf_{bdry}^{B}(-\infty)\,,
\label{fanfbn}
\ee
from (\ref{fafb}).
Thus, adding these two contributions, one has $c_{eff} = 1$ in the UV limit.  
We have plotted the results for $c_{eff}$ versus $\ln mL$ for arbitrarily chosen values of boundary parameters $a_{\pm}$ in Figure 1. It can be seen here that the 
$c_{eff} \rightarrow 1$ as $mL \rightarrow 0$, which agrees with the above UV limit calculations. This feature is obviously independent of the values $a_{\pm}$ assume, 
contrary to the case studied in \cite{AN} (See Eqs. (3.30) and (3.31) of this reference.), where a constraint exists among the boundary parameters $a_{\pm}$.
Moreover, one can also notice the expected $c_{eff} \rightarrow 0$ behaviour at the $mL \rightarrow \infty$ limit.  

\begin{figure}[!h]
	\centering
	\includegraphics[width=0.80\textwidth]{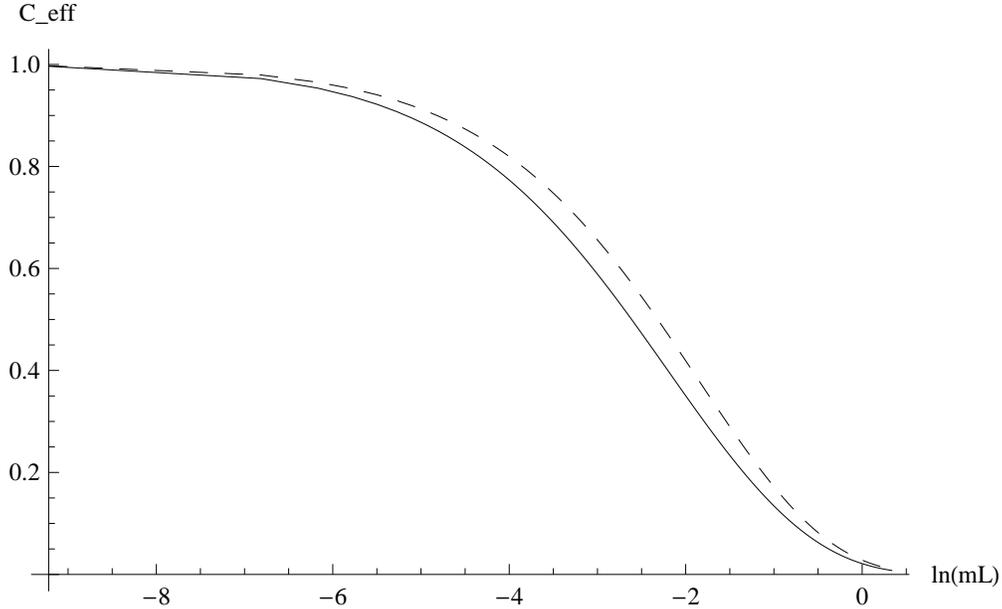}
	\caption[xxx]{\parbox[t]{1.0\textwidth}{
	$c_{eff}$ vs. $\ln mL$, for $a_{+}=1.10, a_{-}= 0.85$ (solid line) and 
	\newline $a_{+}=-0.57, a_{-}= -0.64$ (dashed line)}
	}
\label{fig:graph1}
	\end{figure}

\section{Discussion}\label{sec:conclude}
Utilizing a set of $T-Q$ equations of an open spin-$\frac{1}{2}$ XXZ quantum spin chain with nondiagonal boundary terms, we investigate the
free-Fermion point of a boundary sine-Gordon (bsG) model with nondiagonal boundary interactions. A set of auxiliary functions are derived from these $T-Q$
equations to obtain the Casimir energy for the ground state of the bsG model at the free-Fermion point.
Contrary to a previous case studied where the lattice boundary parameters (and the continuum boundary parameters) obey a certain complex constraint relation (and thereby two real constraints)\cite{AN},
here two of these parameters are set to be completely arbitrary. The Casimir energy for the ground state obtained is then compared and shown to agree with the TBA results obtained in 
\cite{CSS2}. The relations between the lattice parameters $a_{\pm}$ and continuum parameters $\eta_{\pm}$ obtained in \cite{AN} are 
then shown to hold true here as well. This suggests that these relations might hold true in general. From the Casimir energy, we computed the effective central charge of the 
bsG model at the free-Fermion point in the UV limit, which can be seen to agree with the general plots of $c_{eff}$ versus $\ln (mL)$. They also indicate that 
the effective central charge in the UV limit is independent of the boundary parameters, a result that differs from that given in \cite{AN}, for the case where the boundary
parameters obey a certain constraint. This feature where the effective central charge in the UV limit being independent of the boundary parameters is expected for models 
with Neumann boundary condition as is the case here (nondiagonal boundary terms). Interestingly, the case studied in \cite{AN} which is also a model with Neumann boundary 
condition does not have this property for its effective central charge. Perhaps one may attribute this to spectral equivalences between open spin-$\frac{1}{2}$ XXZ chain models 
with diagonal-diagonal and nondiagonal-nondiagonal boundary terms. Spectral equivalences between these open XXZ spin chains have been shown 
to exist \cite{Nichols, deGier, ZoltBaj}. To our knowledge, such equivalences have been found when the boundary parameters obey certain constraint as in \cite{AN}. 
This could potentially explain the seeming contradiction in the effective central charge in the UV limit, where it is independent of the boundary parameters here and 
not in \cite{AN}. 

As part of future work, other possible problems can be explored further. One can extend the analysis in this paper to include values of $\mu$ other than $\pi/2$. This
will result in a set of nonlinear integral equations which would be crucial in the computation of the Casimir energy. Also, numerical work to calculate
the effective central charge (for values of $\mu$ other than $\pi/2$) which include intermediate volume regions is of particular interest. In addition, work carried out here may be extended to the
general case of the nondiagonal boundary interactions using solutions found in \cite{MNC}. Finally, although we have considered only the ground state here,
it would be interesting to investigate the excited states as well. We hope to address some of these issues in the future.  

\section*{Acknowledgments}
I am grateful to the referees for their invaluable suggestions and comments that have helped to greatly improve the presentation of the paper.

\end{document}